\documentclass[fleqn,twoside]{article}
\usepackage{espcrc2}

\usepackage{graphicx}
\usepackage[figuresright]{rotating}
\usepackage{amssymb}
\usepackage{cite}
\newcommand{\AmS}{{\protect\the\textfont2
  A\kern-.1667em\lower.5ex\hbox{M}\kern-.125emS}}

\title{Physics at an Upgraded Fermilab Proton Driver}

\author{S. Geer\address[MCSD] {Fermi National Accelerator Laboratory,\\
        P.O. Box 500, Batavia, Illinois, U.S.A.\\} 
        \thanks{This work was supported by the Fermi National Accelerator 
Laboratory, which is operated by Universities Reasearch Association, under 
contract No.~DE-AC02-76CH03000 with the U.S. Department of Energy.}}
       
\begin{document}

\vspace{1pc}
\begin{abstract}
In 2004 the Fermilab Long Range Planning Committee identified a new high intensity 
Proton Driver as an attractive option for the future, primarily motivated by the 
recent exciting developments in neutrino physics. Over the last few months a 
physics study has developed the physics case for the Fermilab Proton Driver. 
The potential physics opportunities are discussed.
\end{abstract}

\maketitle

\section{INTRODUCTION}

In the last few years there has been interest in a new generation 
of high intensity multi-GeV proton accelerators. 
At Fermilab the design that is currently favored~\cite{foster,pd-website,kephart} 
consists of an 
8~GeV $H^-$ superconducting (SC) Linac that utilizes 
International Linear Collider (ILC) technology. The Linac would  
produce a 0.5~megawatt beam which could be upgraded to 2~megawatts. 
A small fraction of the 8~GeV beam would be used to 
fill the Fermilab Main Injector (MI) with the maximum number of protons 
that, with some modest improvements, it can accelerate. 
This would yield a 2~megawatt MI beam at an 
energy anywhere within the range 40~GeV to 120~GeV.
 
\begin{figure}[b!]
\begin{center}
\scalebox{.35}{\rotatebox{270}{\includegraphics*[bb=78 61 532 721,clip]{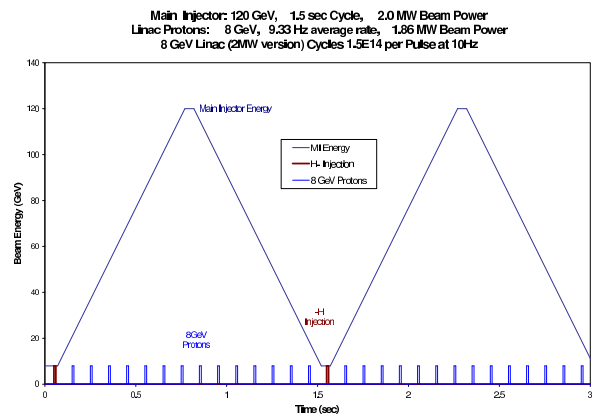}}}
\end{center}
   \caption{\label{fig:MIcycle} Proton Driver bunch structure and the Main Injector cycle.
}
\end{figure}

Hence the upgraded proton source would simultaneously deliver two beams: 
a 2~megawatt beam at MI energies, and an  
$\sim 0.5 - 2$~megawatt beam at 8~GeV. To illustrate this the cycle structure 
is shown in Fig.~\ref{fig:MIcycle}. The MI would receive one pulse from the Linac 
every 1.5~sec. Note that the MI fill time is very short ($<1$~ms).  
The MI cycle time is dominated by the time to ramp up to 120~GeV and ramp down 
to 8~GeV. The 14 Linac pulses that are available, while the MI is ramping 
and at flat top, would provide beam for an 8~GeV program. Improvements in the 
MI ramping time might eventually enable more of the 8~GeV Linac beam to be 
accelerated in the MI, yielding beam powers exceding 2~megawatts.

\section{MOTIVATION}

The interest in a new Fermilab Proton Driver is motivated by the exciting 
discoveries that have been made in the neutrino sector. 
In the last few years solar, atmospheric, and reactor neutrino experiments have 
revolutionized our understanding of the nature of neutrinos. We now know that 
neutrinos produced in a given flavor eigenstate can transform themselves into 
neutrinos of a different flavor as they propagate over macroscopic 
distances. This means that, like quarks, neutrinos have a non-zero mass, the flavor 
eigenstates are different from the mass eigenstates, and hence neutrinos mix. 
However, we have incomplete knowledge of the properties of neutrinos since 
{\it we do not know the spectrum of neutrino masses, and we have only partial 
knowledge of the mixing among the three known 
neutrino flavor eigenstates}.  Furthermore, it is possible that the simplest 
three-flavor mixing scheme is not the whole story, and that a complete 
understanding of neutrino properties will require a more complicated framework. 
In addition to determining the parameters that describe the neutrino sector, 
the three-flavor mixing framework must be tested.

The discovery that neutrinos have mass is exciting. The Standard Model (SM) 
cannot accommodate non-zero neutrino mass terms without some 
modification. We must either introduce right-handed neutrinos (to generate 
Dirac mass terms) or allow neutrinos to be their own antiparticle (violating 
lepton number conservation, and allowing Majorana mass terms). Hence 
{\it the physics of neutrino masses is physics beyond the Standard Model}.
Although we do not know the neutrino mass spectrum, we do know that the masses, 
and the associated mass-splittings, are tiny compared to the masses of any other 
fundamental fermion. This suggests that the physics responsible for neutrino mass 
will include new components radically different from those of the SM. 
Furthermore, although we do not have complete knowledge of the mixing between 
different neutrino flavors, we do know that it is qualitatively very different 
from the corresponding mixing between different quark flavors. The observed 
difference necessarily constrains our ideas 
about the underlying relationship between quarks and leptons, and hence models 
of quark and lepton unification in general, and Grand Unified Theories (GUTs) in 
particular. Note that in neutrino mass models the seesaw mechanism~\cite{Yanagida:1980,
Glashow:1979vf,Gell-Mann:1980vs,Mohapatra:1980ia,Valle} provides a 
quantitative explanation for the observed small neutrino masses, which arise as 
a consequence of the existence of right-handed neutral leptons at the GUT-scale. 
Over the last few years, as our knowledge of the neutrino oscillation parameters 
has improved, a previous generation of neutrino mass models has already been ruled 
out, and a new set of models has emerged specifically designed to accommodate the 
neutrino parameters. Further improvement in our knowledge of the oscillation 
parameters will necessarily reject many of these models, and presumably encourage 
the emergence of new ideas. Hence {\it neutrino physics is experimentally driven, 
and the experiments are already directing our ideas about what lies beyond the 
Standard Model}.

In addition to providing clues about physics beyond the SM,  
understanding neutrino properties is also important 
because neutrinos are the most common matter particles in the universe. In number, 
they exceed the constituents of ordinary matter (electrons, protons, neutrons) 
by a factor of ten billion. They probably account for at least as much energy in the 
universe as all the stars combined and, depending on their exact masses, might 
also account for a few percent of the so-called ``dark matter''. In addition, 
neutrinos are important in stellar processes. There are 
70 billion per second 
streaming through each square centimeter of the Earth from the Sun.  
Neutrinos govern the dynamics of supernovae, and hence the production of heavy 
elements in the universe. Furthermore, if there is CP violation in the neutrino 
sector, the physics of neutrinos in the early universe might ultimately be 
responsible for baryogenesis.
{\it If we are to understand ``why we are here'' and the basic nature of the 
universe in which we live, we must understand the basic properties of the 
neutrino}.

Our desire to understand both the universe in which we live and physics beyond 
the SM provides a compelling case for an experimental program that 
can elucidate the neutrino mass spectrum, measure neutrino mixing, and test the 
three-flavor mixing framework.
To identify the best ways to address the most important open neutrino 
questions, and to determine an effective, fruitful U.S. role within a 
global experimental neutrino program, the American Physical Society's 
Divisions of Nuclear Physics and Particles and Fields, together with the 
Divisions of Astrophysics and the Physics of Beams, have recently conducted 
a ``Study on the Physics of Neutrinos''. This study recommended~\cite{the-neutrino-matrix}
{\it ``... as a high priority, a comprehensive U.S. program to 
complete our understanding of neutrino mixing, to determine the character 
of the neutrino mass spectrum, and to search for CP violation among 
neutrinos'' }, and identified, as a key ingredient of the 
future program,  
{\it ``A proton driver in the megawatt class or above and neutrino superbeam 
with an appropriate very large detector capable of observing CP violation 
and measuring the neutrino mass-squared differences and mixing parameters 
with high precision.'' } The proposed Fermilab Proton Driver would, 
together with a suitable new detector,  
fullfill this need by providing a 2~megawatt proton beam at Main Injector 
(MI) energies for the future NuMI~\cite{numi} program.

The NuMI beam is unique. It is the only neutrino beam that has an appropriate 
energy and a sufficiently long baseline to produce, due to matter effects, 
significant changes in the effective oscillation parameters. These matter 
effects can be exploited to determine the pattern of neutrino masses. 
Furthermore, when combined with measurements from the much-shorter-baseline 
T2K experiment~\cite{t2k} being built in Japan,
an appropriate NuMI-based experiment could exploit matter effects to achieve 
a greatly enhanced sensitivity to CP violation in the neutrino sector. 

Although neutrino oscillations provide the primary motivation for interest in 
the Fermilab Proton Driver, the participation in recent 
proton driver physics workshops has been broader than the neutrino physics 
community. Note that intense neutrino, muon, pion, kaon, neutron, and antiproton 
beams at the Fermilab Proton Driver would offer great flexibility 
and could support a diverse program of experiments of 
interest to particle physicists, nuclear physicists, and nuclear-astrophysicists.
In particular, as the Large Hadron Collider (LHC) and the ILC begin to 
probe the energy frontier, a new round of precision flavor physics experiments 
at the Fermilab Proton Driver 
would provide information that is complementary to the LHC and ILC data by indirectly 
probing high mass scales through radiative corrections. This would help to 
elucidate the nature of any new physics that is discovered at the energy frontier.

\section{OSCILLATION MEASUREMENTS}

To understand the neutrino oscillation physics reach at the Fermilab Proton Driver 
we first introduce the three-flavor mixing parameters, and then discuss event rates 
and discovery potential.

\subsection{Three-Flavor Mixing Parameters}

There are three known neutrino flavor eigenstates 
$\nu_\alpha = (\nu_e, ~\nu_\mu, ~\nu_\tau)$. Since transitions have been 
observed between the flavor eigenstates we now know that neutrinos have 
non-zero masses, and that there is mixing between the flavor eigenstates.
The mass eigenstates $\nu_i = (\nu_1, ~\nu_2, ~\nu_3)$ with masses 
$m_i = (m_1, ~m_2, ~m_3)$ are related to the flavor eigenstates by 
a $3 \times 3$ unitary mixing matrix $U^\nu$~\cite{mns}, 
\begin{equation}
  |\nu_\alpha\rangle = \sum_i ( U^\nu_{\alpha i} )^* |\nu_i\rangle
\label{mix}
\end{equation}
Four numbers are needed to specify all of the matrix elements, namely three mixing 
angles ($\theta_{12}, \theta_{23}, \theta_{13}$) and one complex 
phase ($\delta$). In terms of these parameters:  $ U^\nu =$
\begin{equation}
\left( \begin{array}{ccc}
  c_{13} c_{12}       & c_{13} s_{12}  & s_{13} e^{-i\delta} \\
\\
-c_{23} s_{12}
& c_{23} c_{12}
& c_{13} s_{23} \\
-s_{13} s_{23} c_{12} e^{i\delta}
& -s_{13} s_{23} s_{12} e^{i\delta}
& \\
\\
    s_{23} s_{12}
& -s_{23} c_{12}
& c_{13} c_{23}\\
   -s_{13} c_{23} c_{12} e^{i\delta}
& -s_{13} c_{23} s_{12} e^{i\delta}
& 
\end{array} \right) \,
\label{mns}
\end{equation}
where $c_{jk} \equiv \cos\theta_{jk}$ and $s_{jk} \equiv \sin\theta_{jk}$. 
Neutrino oscillation measurements have already provided some knowledge 
of $U^\nu$, which is approximately given by:
\begin{equation}
U^\nu =
\left( \begin{array}{ccc}
  0.8  & 0.5  & ? \\
  0.4  & 0.6  & 0.7 \\
  0.4  & 0.6  & 0.7 \\
\end{array} \right) \,
\label{mnsnumbers}
\end{equation}
We have limited knowledge of the (1,3)-element of the mixing matrix. This 
matrix element is parametrized by $s_{13} e^{-i\delta}$. We have only an upper 
limit on $\theta_{13}$ and no knowledge of $\delta$. 
Note that $\theta_{13}$ and $\delta$ are particularly important because if 
$\theta_{13}$ and $\sin \delta$ are non-zero there will be CP violation in 
the neutrino sector.

\begin{figure}[t]
\includegraphics[width=0.5\textwidth]{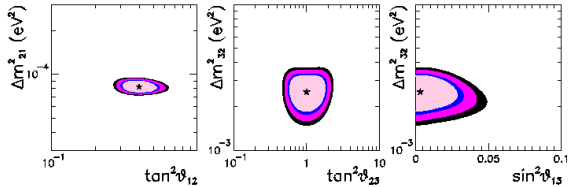}
\caption{\label{parsfig} Current experimental constraints on the three mixing 
angles $\theta_{12}$, $\theta_{23}$, and $\theta_{13}$ and  
on the two mass-squared differences $\Delta m_{12}^2$ and 
$\Delta m_{23}^2$. The star indicates the most likely solution. 
Figure taken from \cite{the-neutrino-matrix}.}
\end{figure}

\begin{figure}[t]
\begin{center}
\includegraphics[width=0.5\textwidth]{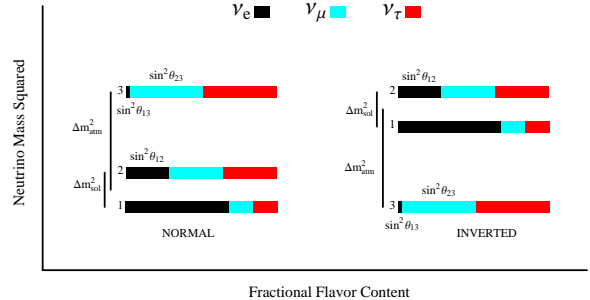}
\end{center}
\caption{\label{hierfig}
The two possible arrangements of the masses of the three known neutrinos, 
based on neutrino oscillation measurements. The spectrum on the left 
corresponds to the {\it Normal Hierarchy} and has $\Delta m^2_{32} > 0$. 
The spectrum on the right corresponds to the {\it Inverted Hierarchy} and 
has $\Delta m^2_{32} < 0$.
The $\nu_e$ fraction of each mass eigenstate is indicated by the black solid
region. The $\nu_\mu$ and $\nu_\tau$ fractions are indicated by
the blue (red) regions respectively.
The $\nu_e$ fraction in the
mass eigenstate labeled ``3'' has been set to the CHOOZ bound. 
Figure from Ref.~\cite{Mena:2004}.}
\end{figure}
Neutrino oscillations are driven by the splittings between the neutrino 
mass eigenstates. 
It is useful to define the differences between the squares of the masses of 
the mass eigenstates $\Delta m^2_{ij} \equiv m^2_i-m^2_j$. 
The probability that a neutrino of energy $E$ and initial flavor $\alpha$ will 
``oscillate'' into a neutrino of flavor $\beta$ is given by  
$P_{\alpha \beta} \equiv P(\nu_\alpha \rightarrow \nu_\beta) =
\left| \langle \nu_\beta | \exp( - i \mathcal{H} t ) | \nu_\alpha \rangle \right|^2$,
which in vacuum is given by 
\begin{equation}
P_{\alpha \beta}  =  
\sum\limits_{j=1}^3 \, \sum\limits_{k=1}^3 U_{\alpha j} U_{\alpha k}^* U_{\beta j}^* U_{\beta k} \, 
\exp \left( - i \frac{\Delta m_{kj}^2}{2 E} t \right)
\label{equ:oscgeneral}
\end{equation}
If neutrinos of energy $E$ travel a distance $L$ then a 
non-zero $\Delta m^2_{ij}$ will result in neutrino flavor 
oscillations that have maxima at given values of $L/E$, and oscillation 
amplitudes that are determined by the matrix elements $U^\nu_{\alpha i}$, and 
hence by $\theta_{12}, \theta_{23}, \theta_{13}$, and $\delta$. 

Our present knowledge of the neutrino mass splittings and mixing matrix, has been 
obtained from atmospheric \cite{Fukuda:2000np,Fukuda:1998mi}, solar~\cite{Lande:2003ex,
Abdurashitov:1999bv,Hampel:1998xg,Ahmad:2001an,Ahmed:2003kj,Fukuda:2002pe}, 
reactor \cite{Apollonio:1999ae,Boehm:2001ik,Eguchi:2002dm}, and 
accelerator-based \cite{Aliu:2004sq} neutrino experiments, and is summarized in 
Fig.~\ref{parsfig}. The 
solar-neutrino experiments and the reactor experiment KamLAND probe values of 
$L/E$ that are sensitive to $\Delta m^2_{21}$, and the mixing angle $\theta_{12}$. 
Our knowledge of these parameters is shown in the left panel of Fig.~\ref{parsfig}.
The atmospheric-neutrino experiments and the accelerator based experiment K2K probe 
values of $L/E$ that are sensitive to $\Delta m^2_{32}$, and the mixing angle 
$\theta_{23}$. Our knowledge of these parameters is shown in the central panel of Fig.~\ref{parsfig}.
Searches for $\nu_\mu \leftrightarrow \nu_e$ transitions with values of $L/E$ corresponding
to the atmospheric-neutrino scale are sensitive to the third mixing angle $\theta_{13}$. 
To date these searches have not observed this transition, and hence we have only an 
upper limit on $\theta_{13}$, which comes predominantly from the CHOOZ reactor 
experiment~\cite{Apollonio:1999ae}, and is shown in the right panel of Fig.~\ref{parsfig}.

\begin{table*}[htb]
\caption{Signal and background $\nu_\mu \rightarrow \nu_e$ event rates
for values of $\theta_{13}$ that are just below the present upper limit 
and an order of magnitude below the upper limit. The rates are for 
the normal mass hierarchy and $\delta = 0$. The numbers for each 
experiment correspond to 5~years of running with the nominal beam intensities.
}
\label{table:1}
\newcommand{\m}{\hphantom{$-$}}
\newcommand{\cc}[1]{\multicolumn{1}{c}{#1}}
\renewcommand{\tabcolsep}{2pc} 
\renewcommand{\arraystretch}{1.2} 
\begin{tabular}{@{}llll}
\hline
Experiment           & Signal & Signal & Background \\
                     & $\sin^2 2\theta_{13}=0.1$ & $\sin^2 2\theta_{13}=0.01$ & \\
\hline
MINOS  & 49.1 & 6.7 & 108\\
ICARUS & 31.8 & 4.5 & 69.1\\
OPERA  & 11.2 & 1.6 & 28.3\\
T2K    & 132  & 16.9 & 22.7\\
NO$\nu$A& 186&23.0&19.7\\
NO$\nu$A$+$FPD&716&88.6&75.6\\
NuFACT (neutrinos)&29752&4071&44.9\\
NuFACT (antineutrinos)&7737&1116&82.0\\
\hline
\end{tabular}\\[2pt]
From the calculations of W.~Winter, based on the Globes program~\cite{Huber:2002mx}.
\end{table*}
\begin{table*}[htb]
\caption{Signal and background $\nu_\mu \rightarrow \nu_e$ event rates
for $\theta_{13}$ just below the present upper limit 
($\sin^2 2 \theta_{13} = 0.1$). The rates are for 
the normal and inverted mass hierarchies with $\delta = 0$ (no CP violation) 
and $\delta = \pi/2$ (maximal CP violation). The numbers for each 
experiment correspond to 5~years of running with the nominal beam intensities.
}
\label{table:2}
\newcommand{\m}{\hphantom{$-$}}
\newcommand{\cc}[1]{\multicolumn{1}{c}{#1}}
\renewcommand{\tabcolsep}{1pc} 
\renewcommand{\arraystretch}{1.2} 
\begin{tabular}{@{}llllll}
\hline
Experiment           & Normal & Normal & Inverted & Inverted & Back- \\
                     &$\delta=0$&$\delta=\pi/2$&$\delta=0$&$\delta=\pi/2$&ground \\
\hline
T2K    & 132  & 96 & 102& 83& 22.7\\
NO$\nu$A& 186&138&111&85&19.7\\
NO$\nu$A$+$FPD&716&531&430&326&75.6\\
NuFACT ($\nu$)&29752&27449&13060&17562&44.9\\
NuFACT ($\overline{\nu}$)&7737&5942&9336&10251&82.0\\
\hline
\end{tabular}\\[2pt]
From the calculations of W.~Winter, based on the Globes program~\cite{Huber:2002mx}.
\end{table*}

The mixing angles tell us about the flavor content of the neutrino mass eigenstates. 
Our knowledge of the $\Delta m^2_{ij}$ and the flavor content of the mass eigenstates 
is summarized in Fig.~\ref{hierfig}. Note that there are two possible patterns of 
neutrino mass. 
This is because the neutrino oscillation experiments to date have been sensitive 
to the magnitude of $\Delta m^2_{32}$, but not its sign. The neutrino spectrum shown 
on the left in Fig.~\ref{hierfig} is called the {\it Normal Mass Hierarchy} and 
corresponds to $\Delta m^2_{32} > 0$.  The neutrino spectrum shown 
on the right is called the {\it Inverted Mass Hierarchy} and corresponds to 
$\Delta m^2_{32} < 0$. The reason we don't know the sign of $\Delta m^2_{32}$, and 
hence the neutrino mass hierarchy, is that neutrino oscillations in vacuum depend 
only on the magnitude of $\Delta m^2_{32}$. However, in matter the effective parameters describing 
neutrino transitions involving electron-type neutrinos are modified~\cite{wolfenstein} 
in a way that is 
sensitive to the sign of $\Delta m^2_{32}$. An experiment with a sufficiently long 
baseline in matter and an appropriate $L/E$ can therefore determine the neutrino 
mass hierarchy. 

Finally, it should be noted that there is a possible complication to the simple 
three-flavor neutrino oscillation picture. 
The LSND~\cite{Aguilar:2001ty} experiment  has reported evidence for muon 
anti-neutrino to electron anti-neutrino transitions for values of $L/E$ which 
are two orders of magnitude smaller than the corresponding values observed for 
atmospheric neutrinos. The associated transition probability is very small, 
of the order of 0.3\%.
If this result is confirmed by the MiniBooNE~\cite{Church:1997ry} 
experiment, it will require a third characteristic $L/E$ range for neutrino 
flavor transitions. Since each $L/E$ range implies a different mass-splitting 
between the participating neutrino mass eigenstates, confirmation of the LSND 
result would require more than three mass eigenstates. This would be an exciting 
and radical development. Independent of whether the LSND result is confirmed 
or not, it is important that the future global neutrino oscillation program is 
able to make further tests of the three-flavor oscillation framework.

\subsection{Event Rates}

To obtain sufficient statistical sensitivity to determine the pattern of 
neutrino masses and search for CP violation over a large region 
of parameter-space will require a new detector with a fiducial mass of tens
of kilotons and a neutrino beam with the highest practical intensity. 
To illustrate this, consider the NuMI event rates in the far detector. 
The present NuMI primary proton beam intensity is roughly $10^{13}$~protons 
per second at 120~GeV, which corresponds to 0.2~megawatts on target. 
These protons are used to make a secondary charged pion beam, which is 
focussed into a parallel beam using magnetic horns. The pion beam is 
then allowed to decay whilst propagating down a long decay channel, 
to create a tertiary beam of muon-neutrinos. At the far detector, 
735~km downstream of the target, there are $10^{-5}$ neutrino interactions 
in a 1~kt detector for every $10^{13}$ protons on target. Note that we 
are interested in $\nu_\mu \rightarrow \nu_e$ oscillations, and that 
the present upper limit on $\theta_{13}$ implies that the relevent 
oscillation amplitude is at most $\sim5$\%. Putting these numbers 
together one quickly concludes that we will need proton beam powers 
of one or a few megawatts together with detectors of a few times 
10~kt.

\begin{figure}[!htb]
   \centering \includegraphics[width=0.5\textwidth]{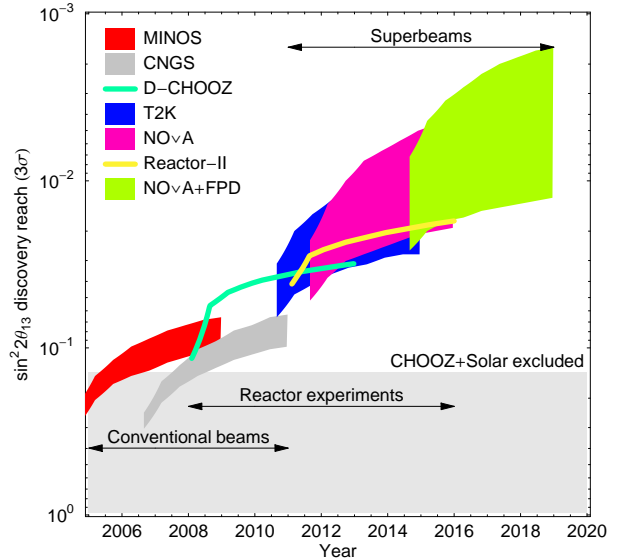}
   \vspace*{0.1cm}
   \caption{ \label{fig:timescale_small} Anticipated evolution of the $\theta_{13}$ discovery 
reach. The 3$\sigma$ sensitivities for the observation of 
a non-zero $\sin^2 2\theta_{13}$. The bands reflect the 
dependence on the CP phase $\delta$. The calculations are for a normal mass hierarchy and are 
based on the simulations in ~\cite{Huber:2002mx,Huber:2004ug} and include 
statistical and systematic uncertainties and parameter correlations. 
All experiments are operated with neutrino running only. 
The starting times of the experiments correspond to those 
stated in the respective LOIs. ReactorII and FPD refer, respectively, to a second 
generation reactor experiment and to the Fermilab Proton Driver.
   }
\end{figure}
To be explicit, the expected $\nu_\mu \rightarrow \nu_e$ event rates 
for future experiments are listed in Table~\ref{table:1} for two values 
of $\theta_{13}$. Note that signal and background rates for the T2K and 
NO$\nu$A~\cite{nova} experiments are comparable and will at best (for the most 
favorable $\theta_{13}$) yield data samples of only $\sim 100$~events. 
This may be sufficient to pin down $\theta_{13}$, is at best barely 
sufficient to make the first determination of the mass hierarchy, and is 
inadequate to search for CP violation or make precision measurements of 
the interesting parameters. The Fermilab Proton Driver, together with 
NO$\nu$A, does significantly better.

Although the signal and background rates for T2K and NO$\nu$A are 
comparable, the two experiments are complementary in the way they 
are sensitive to the oscillation parameters. This is illustrated 
in Table~\ref{table:2} which compares event rates for the two 
mass hierarchies and for CP conserving and CP violating values of $\delta$.
With the Proton Driver, there is a statistically significant 
dependence of event rates on both the mass hierarchy and the phase 
$\delta$. In contrast, the T2K rates are not as sensitive to the 
mass hierarchy. Hence the combination of both experiments provides a  
way to disentangle the parameters. 

\subsection{Discovery Potential}

To complete our knowledge of the neutrino mixing matrix and the pattern 
of neutrino masses we must measure $\theta_{13}$ and $\delta$, determine the sign of 
$\Delta m^2_{32}$, and test the three-flavor mixing framework. The initial 
goal for a Fermilab Proton Driver experiment will be to make these measurements. How far 
this physics program can be pursued will depend upon the magnitude of the unknown mixing 
agle $\theta_{13}$.

The anticipated evolution of the $\sin^2 2\theta_{13}$ discovery reach of the 
global neutrino oscillation program is illustrated in Fig.~\ref{fig:timescale_small}. 
The sensitivity is expected to improve by about an order of magnitude over the 
next decade. This progress is likely to be accomplished in several steps, 
each yielding a factor of a few increased sensitivity. 
During this first decade the Fermilab program will have contributed to the 
improving global sensitivity with MINOS, followed by NO$\nu$A. MINOS is the on-ramp 
for the US long-baseline neutrino oscillation program. NO$\nu$A would be the next step. 
Note that we assume that NO$\nu$A starts taking data with the existing beamline before 
the Proton Driver era. The Proton Driver would take NO$\nu$A into the fast lane of the 
global program. Also note that the accelerator based and reactor based experiments are 
complementary. In particular, the reactor experiments make disappearance measurements, 
limited by systematic uncertainties. The NO$\nu$A experiment is an appearance experiment, 
limited by statistical uncertainties, and probes regions of parameter space beyond the 
reach of the proposed reactor experiments.
\begin{figure*}[t]
\includegraphics[width=\textwidth]{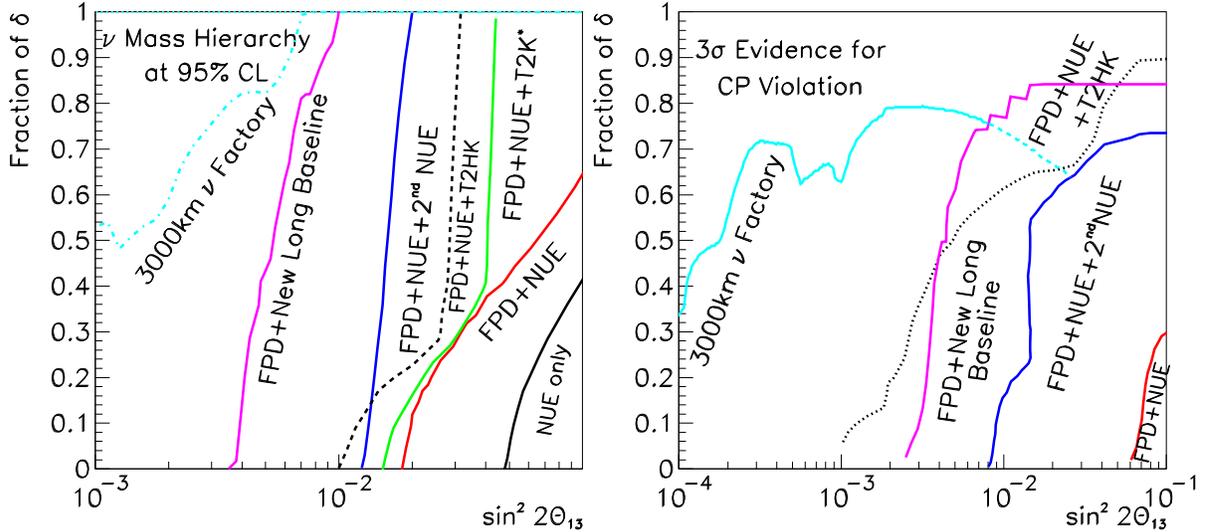}
\caption{
Regions of parameter space where the mass hierarchy (left) and CP 
violation (right) can be observed at 95\% CL and at 3$\sigma$, 
respectively. The label NUE refers to the NO$\nu$A experiment, 
and FPD to the Fermilab Proton Driver. 
T2K$^\star$ refers to an upgraded T2K experiment with 
a 4~megawatt primary beam.
\label{fig:reach_all_pd}}
\end{figure*}

Although we don't know the value of $\theta_{13}$ we have no reason to suspect 
that it is very small. Hence, any of the experiments on the trajectory show in 
Fig.~\ref{fig:timescale_small} might establish a finite value for $\theta_{13}$. 
At this point the focus of the experimental program will change from establishing 
the magnitude of $\theta_{13}$ to measuring the mass hierarchy and searching for 
CP violation. 
Independent of the value of $\theta_{13}$ the initial Fermilab Proton 
Driver long-baseline neutrino experiment (NO$\nu$A+FPD) would be expected to make an 
important contribution to the global oscillation program. 
If $\theta_{13}$ is very small NO$\nu$A+FPD would provide the 
most stringent limit on this important parameter, and prepare the 
way for a neutrino factory~\cite{nf}. If $\theta_{13}$ is sufficiently large, 
NO$\nu$A+FPD would be expected to measure its value, perhaps determine the mass 
hierarchy, and prepare the way for a sensitive search for CP violation.  
The evolution of the Fermilab Proton Driver physics program beyond the 
initial experiments will depend on the value of $\theta_{13}$ and on what other 
neutrino experiments are built elsewhere in the world. 
Hence, in considering the long-term evolution 
of the Fermilab Proton Driver program we must take into account the 
uncertainty on the magnitude of $\theta_{13}$ and consider how the 
global program might evolve.

The experiments needed to determine the mass hierarchy and discover (or place 
stringent limits on) CP violation will depend upon both $\theta_{13}$ and on 
$\delta$. The fractions of all possible values of $\delta$ for which a discovery 
can be made are shown as a function of $\sin^2 2\theta_{13}$  in 
Fig.~\ref{fig:reach_all_pd} for various experiments. The left panel shows the 
potential for determining the mass hierarchy and the right panel for making a 
first observation of CP violation. Note that without a megawatt-class proton source 
none of the future experiments will be able to make a sensitive search for 
CP violation. The NO$\nu$A experiment (labelled NUE in the figure) can make a 
first determination of the mass hierarchy, but only over a very limited region 
of parameter space. The Fermilab Proton Driver would significantly improve the 
prospects for determining the mass hierarchy, and if $\theta_{13}$ is relatively 
large, would enable the first sensitive search for CP violation. Combining NO$\nu$A 
and T2K results would enable further progress if the T2K experiment was upgraded 
to achieve a factor of a few larger data samples (T2K$^\star$). The mass hierarchy 
could then be determined independent of $\delta$ provided $\sin^2 2\theta_{13}$ 
exceeds about 0.04. Smaller values of $\theta_{13}$ will motivate a much more 
ambitious experimental program which will probably include a Neutrino Factory. 
Larger values of $\theta_{13}$ will still motivate a more ambitious experimental 
program focussed on the precision measurements that would put the presently viable
theoretical models under pressure. The second generation of Fermilab Proton Driver 
experiments might, in this case, include a second off-axis detector and/or a new 
longer-baseline beam. Note that, since a Proton Driver can be used to drive 
a Neutrino Factory, the Fermilab Proton Driver offers great flexibility for a 
second generation program independent of the value of $\theta_{13}$.

\section{OTHER PHYSICS}

\begin{figure*}[t!]
     \centering \includegraphics[width=\textwidth]{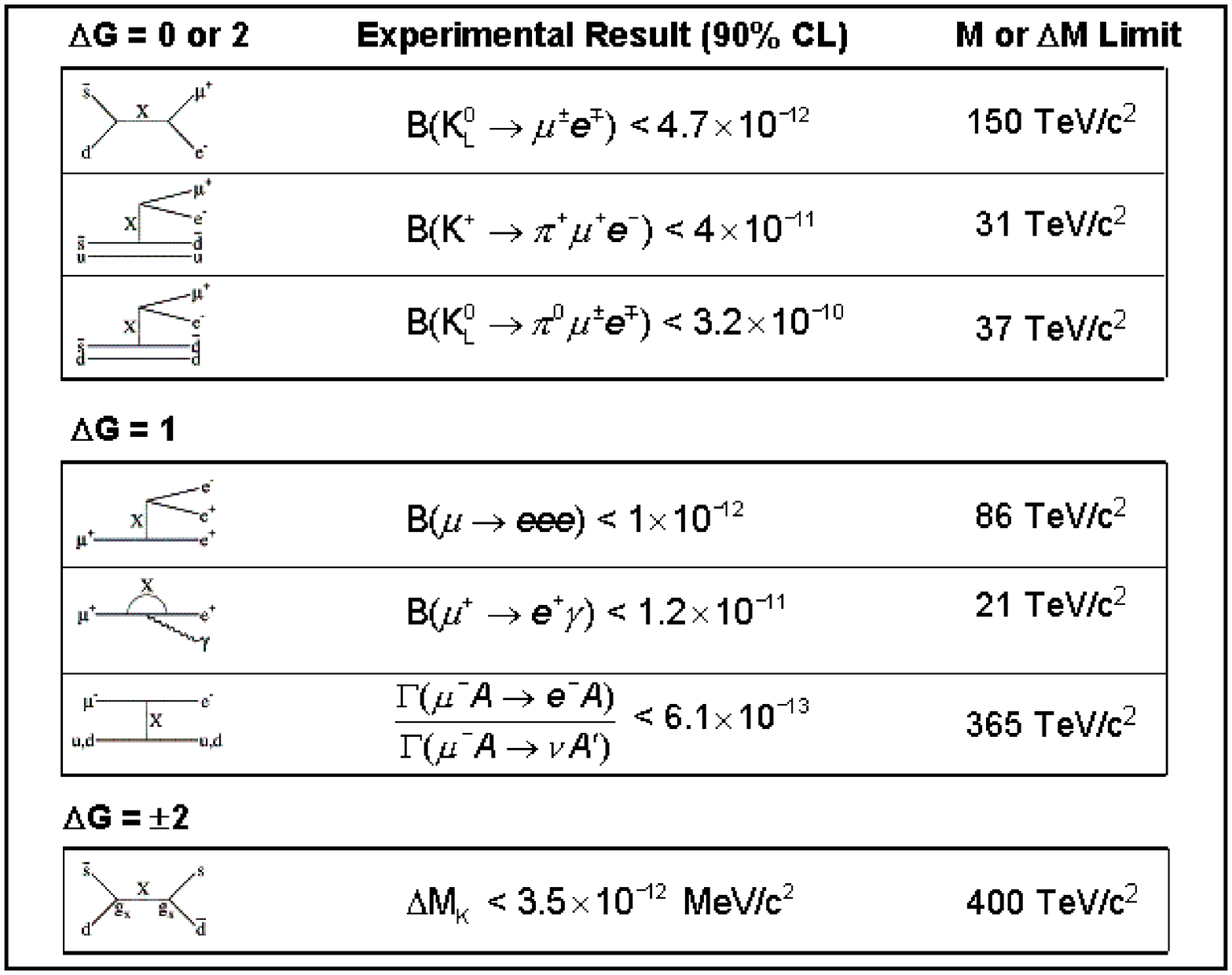}
     \caption {Current limits on Lepton Flavor Violating processes and 
	the mass scales probed by each process.  The upper box is for 
	kaon decays, which involve a change of both quark flavor and 
	lepton flavor.  The bottom box is for muon decays, which involve 
	only lepton flavor change.  The lower limit on the mass scale 
	is calculated assuming the electroweak coupling strength. 
\label{fig:muonkaon}}
  \end{figure*}

 In the past, high precision measurements at low energies have complemented the 
  experimental program at the energy frontier. These low energy experiments not 
  only probe mass scales that are often beyond the reach of colliders, but also 
  provide complementary information at mass scales within reach of the energy 
  frontier experiments. Examples of low energy experiments that have played an 
  important role in this way are muon $(g-2)$ measurements, searches for muon and kaon 
  decays beyond those predicted by the SM, and other measurements of rare muon and kaon 
  processes. A summary of the sensitivity achieved by a selection of these 
  experiments is given in Fig.~\ref{fig:muonkaon}. 
    It seems likely that these types of experiment will continue to 
  have a critical role as the energy frontier moves into the LHC and ILC era. In particular, 
  if the LHC and/or ILC discover new physics beyond the SM, the measurement of 
  quantum corrections that manifest themselves in low energy experiments would be 
  expected to help elucidate the nature of the new physics. If no new physics is 
  discovered at the LHC then precision low energy experiments may provide the only 
  practical way of advancing the energy frontier beyond the LHC in the foreseeable 
  future.

In addition to complementing collider experiments at the energy frontier, 
intense neutrino, muon, pion, kaon, neutron, and antiproton 
beams at the Fermilab Proton Driver could also support a diverse program of 
experiments of interest to particle physicists, nuclear physicists, and 
nuclear-astrophysicists, and offer great flexibility for the future program.
Of the various possibilities that have been considered, neutrino scattering 
physics and the potential physics program that could be pursued with an 
intense low energy muon source offer particularly attractive options that 
would complement, and could be run in parallel with, the neutrino 
oscillation program.

\subsection{Neutrino Scattering}

While neutrino oscillation experiments probe the physics of neutrino masses 
and mixing, neutrino scattering experiments probe the interactions of 
neutrinos with ordinary matter, and enable a search for exotic neutrino 
properties. A complete knowledge of the role of 
neutrinos in the Universe in which we live requires a detailed knowledge 
of neutrino masses, mixing, and interactions. 

Our present knowledge of the neutrino and anti-neutrino scattering 
cross sections in matter is limited. The next generation of approved 
neutrino scattering experiments, including MINER$\nu$A \cite{debbie}
in the NuMI beamline and MiniBooNE \cite{Church:1997ry} using neutrinos 
from the Fermilab Booster, are expected to greatly improve our knowledge. 
In particular, within the next few years we anticipate that precise 
measurements will be made of neutrino scattering on nuclear targets. 
However, we will still lack precise measurements of: 
\begin{itemize}
  \item Anti-Neutrino scattering on nuclear targets.
  \item Neutrino and anti-neutrino scattering on nucleon (hydrogen and deuterium) targets.
  \item Neutrino-electron scattering.
\end{itemize}
The anti-neutrino rates per primary proton on target are, depending on energy, a factor 
of 3-5 less than the neutrino rates. The interaction rates on nucleon 
targets are an order of magnitude less than the corresponding rates on 
nuclear targets, and the cross-section for neutrino-electron scattering is 
considerably smaller than that on nucleons. Hence, beyond the presently approved 
program, a factor of 10-100 increase in data rates will be required to complete the 
neutrino and anti-neutrino scattering measurements. The physics topics that 
could be pursued with a neutrino scattering program at the Fermilab Proton Driver 
include (i) a study of neutrino-electron scattering and a search for a neutrino 
magnetic moment, (ii) measurements of parton distribution functions, particularly at 
large x, (iii) a study of generalized parton distribution functions to determine 
the partonic spatial distributions as a function of longitudinal momenta, 
(iv) measurements of the strange-quark content and spin structure of the nucleon, 
(v) a measurement of the nuclear weak form factor to better understand the 
$G_E$ and $G_M$ measurements at JLab~\cite{jlab}, (vi) studies of 
duality and resonance production to better understand the 
transition between the domain where partons are the appropriate 
degrees of freedom to the domain where baryons and mesons provide the appropriate 
description, and (vii) strange particle production studies to test theoretical 
models~\cite{shrock} of NC induced strange particle production.

In addition to being of interest in their own right, neutrino scattering experiments 
will also play a key role in allowing future precision
oscillation experiments to reach their ultimate sensitivity.  To obtain the most 
precise value of $\Delta m^2_{32}$ (which is eventually required to extract mixing 
angles and the CP-violating phase) we must better understand and quantify the 
nuclear processes interposed between the interaction of an incoming neutrino and 
measurement of outgoing particles in the detector.   Extracting mixing parameters
such as $\theta_{13}$, and ultimately the neutrino mass hierarchy and
CP phase, also requires much better understanding of the neutral current resonant 
and coherent cross-sections that contribute to the background. 
The precision measurement of nuclear effects and exclusive cross-sections will 
provide the necessary foundation for the study of neutrino oscillations with 
high-intensity beams at the Fermilab Proton Driver.

\begin{table*}[t]
\caption{A comparison of the present or near future sensitivities for some representative 
muon experiments with the sensitivities that are in principle attainable with a 
Fermilab Proton Driver.}
\label{tab:muon_sens}
\newcommand{\m}{\hphantom{$-$}}
\newcommand{\cc}[1]{\multicolumn{1}{c}{#1}}
\renewcommand{\tabcolsep}{2pc} 
\renewcommand{\arraystretch}{1.2} 
\begin{tabular}{|c|r|r|} \hline
                & \multicolumn{2}{|c|}{Sensitivity}                             \\
Measurement     & Present or Near Future         & Fermilab Proton Driver       \\ \hline
 EDM $d_\mu$    &  $< 3.7 \times 10^{-19}$ e-cm  & $< 10^{-24} - 10^{-26}$ e-cm \\
  $(g-2)$ $\sigma(a_\mu)$ & $0.2 - 0.5$ ppm      &      0.02 ppm                \\
  BR($\mu\rightarrow e\gamma$) &  $\sim 10^{-14}$ &      $\sim 10^{-16}$         \\
  $\mu A \rightarrow e A$ Ratio & $\sim 10^{-17}$         &      $\sim 10^{-19}$ \\ 
\hline
\end{tabular}
\end{table*}
\subsection{Muon Physics}

Solar-, atmospheric-, and reactor-neutrino experiments have established 
Lepton Flavor Violation (LFV) in the neutrino sector, which suggests the 
existence of LFV processes at high mass scales. Depending on its nature, this 
new physics might also produce observable effects in rare muon processes. 
Furthermore, CP violation in the charged lepton sector, revealed for example 
by the observation of a finite muon Electric Dipole Moment (EDM), might be part 
of a broader baryogenesis via leptogenesis picture. Hence, the neutrino 
oscillation discovery enhances the motivation for a continuing program of 
precision muon experiments. 
In addition, the expectation that there is new physics at the TeV scale also 
motivates a new round of precision muon experiments. LFV muon decays and the muon 
anomalous magnetic moment $a_\mu = (g-2)/2$ and EDM are sensitive probes of new 
dynamics at the TeV scale. In general, with sufficient sensitivity, these 
experiments would help elucidate the nature of new physics observed at the 
LHC and ILC. 

Low energy high precision muon experiments require high intensity beams. 
Since most of the 8~GeV Fermilab Proton Driver beam from the SC linac would not be 
used to fill the MI, it would be available to drive a high intensity muon source. 
In addition to high intensity, precision muon experiments also require an 
appropriate bunch structure, which varies with experiment.
In the post-collider period it might be possible to utilize the Recycler Ring to 
repackage the 8~GeV proton beam, yielding a bunch structure optimized for each 
experiment. The combination of Proton Driver plus Recycler Ring would provide the front-end 
for a unique muon source with intensity and flexibility that exceed any existing 
facility.

The Recycler is an 8 GeV storage ring in the MI tunnel that can run at the same 
time as the MI. The beam from the Fermilab Proton Driver SC linac that is 
not used to fill the MI could be used to fill the Recycler Ring approximately ten 
times per second.  The ring would then be emptied gradually in the 100~ms intervals 
between linac pulses. Extraction could be continuous or in bursts.
For example, the Recycler Ring could be loaded with one linac pulse of 
$1.5 \times 10^{14}$ protons every 100~ms, with one missing pulse every 1.5 seconds 
for the 120 GeV MI program.  This provides $\sim 1.4 \times 10^{22}$ protons at 
8~GeV per operational year 
($10^7$~seconds). In the Recycler each pulse of $1.5 \times 10^{14}$ protons can 
be chopped into 588 bunches of $0.25 \times 10^{12}$ protons/bunch with a pulse 
width of 3~ns.  A fast kicker would permit the extraction of one bunch at a time. 
The beam structure made possible by the Proton Driver linac and the Recycler Ring 
is perfect for $\mu \rightarrow$ e conversion experiments, muon EDM searches and 
other muon experiments where a pulsed beam is required.  Slow extraction from the 
Recycler Ring for $\mu \rightarrow e \gamma$ and $\mu \rightarrow 3e$ searches is 
also possible. 

Using an 8~GeV primary proton beam together with a suitable target and 
solenoidal capture and decay channel, the calculated yield of low energy 
muons is $\sim 0.2$ of each sign per incident proton~\cite{brice}. 
With $1.4 \times 10^{22}$ 
protons at 8~GeV per operational year (corresponding to $\sim2$ megawatts) 
this would yield $\sim 3 \times 10^{21}$ muons per year. This muon flux greatly 
exceeds the flux required to make progress in a broad range of muon experiments. 
However, the muons at the end of the decay channel have low energy, a large momentum 
spread, and occupy a large transverse phase space. Without further manipulation their 
utilization will be very inefficient. The interface 
between the decay channel and each candidate experiment has yet to be designed. In 
Japan a Phase Rotated Intense Slow Muon Source 
(PRISM~\cite{prism}) based on an FFAG ring that 
reduces the muon energy spread (phase rotates) is being designed. This phase 
rotation ring has a very large transverse acceptance ($800\pi$~mm-mrad) and a momentum 
acceptance of $\pm30\%$ centered at 500~MeV/c. PRISM reduces the momentum and momentum 
spread to 68~MeV/c and $\pm 1-2\%$ respectively. Hence, a PRISM-like ring 
downstream of 
the decay channel might accept a significant fraction of the muon spectrum 
and provide a relatively  efficient way to use the available muon flux. Explicit 
design work must be done to verify this, but it should be noted that a muon 
selection system that utilizes only $1\%$ of the muons available at the end of the 
decay channel will still produce an adequate muon flux for most of the desired 
cutting-edge experiments.
Scaling from  proposals for muon experiments at JPARC, and making some plausible 
assumptions about the evolution of detector technology in the coming decade, 
the sensitivities that might be obtained at a Fermilab Proton Driver muon 
source are summarized in Table~\ref{tab:muon_sens} for the leading desired experiments.
Orders of magnitude improvements in sensitivity beyond those already 
acheived would be possible.

Finally, a new 8~GeV multi-megawatt Proton Driver at Fermilab, together with an 
appropriate target, pion capture system, decay channel, and phase rotation 
system could provide the first step toward a Neutrino Factory based on a 
muon storage ring. Hence, the development of a cutting edge muon program at the 
Fermilab Proton Driver is a particularly attractive complement to the 
long-term neutrino oscillation program.

\section{CONCLUSIONS}

In 2004 the Fermilab Long Range Planning Committee~\cite{flrp} 
identified a new high intensity 
Proton Driver as an attractive option for the future, primarily motivated by the 
recent exciting developments in neutrino physics. Over the last few months a 
physics study~\cite{pd-website} has developed the physics case for the Fermilab 
Proton Driver that is described in this paper. The conclusions from the study are:

\begin{enumerate}
\item Independent of the value of the unknown mixing angle $\theta_{13}$ 
an initial Fermilab Proton Driver long-baseline neutrino experiment 
will make a critical contribution to the global oscillation program.
\item  If $\theta_{13}$ is very small the initial Fermilab Proton Driver 
experiment will provide the most stringent limit on 
$\theta_{13}$ and prepare the way for a neutrino factory. The 
expected $\theta_{13}$ sensitivity exceeds that expected for reactor-based 
experiments, or any other accelerator-based experiments.  
\item If $\theta_{13}$ is sufficiently large the initial Fermilab Proton 
Driver experiment will precisely measure its value, 
perhaps determine the mass hierarchy, and prepare the way for a sensitive 
search for CP violation. The value of $\theta_{13}$ will guide the 
further evolution of the Proton Driver neutrino program. 
\item The Fermilab Proton Driver neutrino experiments will also make 
precision measurements of the other oscillation parameters, and conduct 
an extensive set of neutrino scattering measurements, some of which are 
important for the oscillation program. Note that the neutrino scattering 
measurements require the highest achievable intensities at both MI energies and 
at 8~GeV.
\item The Fermilab Proton Driver could also support a broad range of other 
experiments of interest to particle physicists, nuclear physicists, and 
nuclear astrophysicists. These experiments could exploit antiproton- 
and kaon-beams from the MI, or muon-, pion-, or neutron-beams from 
the 8~GeV linac. These ``low energy'' experiments would provide 
sensitivity to new physics at high mass scales which would be complementary to 
measurements at the LHC and beyond.
\end{enumerate}

{\bf Acknowledgements}
\newline
The results summarized in this paper are from the Fermilab Proton Driver 
Physics Study, and therefore include contibutions from all those that 
have participated. I am particularly indebted to the Working Group 
Conveners who organized and pushed forward the study: D.~Harris, 
S.~Brice, W.~Winter, J.~Morfin, R.~Ransome, R.~Tayloe, R.~Ray, 
L.~Roberts, H.~Nguyen, T.~Yamanaka, D.~Christian, M.~Mandlekern, 
H.~Cheung, P.~Kasper, P.~Ratoff, T.~Bowles, and G.~Green.


\begin{thebibliography}{9}

\bibitem{foster}
G. W. Foster and J. MacLachlan, 
``A Multi-Mission 8 GeV Injector Linac as a Fermilab Booster
Replacement'', 
Linac 2002, 
http://accelconf.web.cern.ch/AccelConf/l02/
PAPERS/FR202.PDF

\bibitem{pd-website}
http://protondriver.fnal.gov/

\bibitem{kephart}
R.~Kephart,
``Plans for a Proton Driver at Fermilab'',
Nucl. Phys. B (Proc. Suppl.) 147 (2005) 124-127.

\bibitem{Yanagida:1980}
T. Yanagida,    
"Proceedings of the Workshop on the Unified Theory and
the Baryon Number in the Universe",
KEK, Tsukuba, Japan,
1979.

\bibitem{Glashow:1979vf}
S. L. Glashow,
"THE FUTURE OF ELEMENTARY PARTICLE PHYSICS",
"Proceedings of the 1979 Carg{\`e}se Summer Institute
on Quarks and Leptons",687-713,
1980.

\bibitem{Gell-Mann:1980vs}
M. Gell-Mann, P. Ramond, and R. Slansky,
"COMPLEX SPINORS AND UNIFIED THEORIES",
Supergravity, North Holland, Amsterdam (publisher),
1979.

\bibitem{Mohapatra:1980ia}
R.N. Mohapatra and G. Senjanovi,
"NEUTRINO MASS AND SPONTANEOUS PARITY VIOLATION",
Phys. Rev. Lett. 44 (1980) 912.

\bibitem{Valle}
J. Schechter and J.W.F. Valle, 
Phys. Rev. D22 (1980) 2227; Phys. Rev. D25 (1982) 774.

\bibitem{the-neutrino-matrix}
"The Neutrino Matrix",
Report from the APS DNP/DPF/DPB Joint Study on the Future of Neutrino Physics,
November, 2004.

\bibitem{numi}
Fermilab NuMI Group, 
"NuMI Facility Technical Design Report", 
Fermilab Report NuMI-346, October 1998.

\bibitem{t2k}
Y. Itow et al. 
``The JHF-Kamioka neutrino project'', 
KEK Report 2001-4, ICRR-Report-477-2001-7, TRI-PP-01-05, 
hep-ex/0106019.

\bibitem{mns}
Z. Maki, M. Nakagawa, and S. Sakata,
Prog. Theor. Phys. 28 (1962) 870.

\bibitem{Fukuda:2000np}
S. Fukuda et al. (Super-Kamiokande Collab.),
"Tau neutrinos favored over sterile neutrinos in atmospheric
muon  neutrino oscillations",
Phys. Rev. Lett. 85 (2000) 3999-4003"; 
hep-ex/0009001.

\bibitem{Fukuda:1998mi}
S. Fukuda et al. (Super-Kamiokande Collab.),
"Evidence for oscillation of atmoFukuda:1998mispheric neutrinos",
Phys. Rev. Lett. 81 (1998) 1562-1567; hep-ex/9807003.

\bibitem{Lande:2003ex}
K. Lande and P. Wildenhain,
"The Homestake chlorine solar neutrino experiment: Past, 
present and future",
Nucl. Phys. Proc. Suppl. 118 (2003) 49-54.

\bibitem{Abdurashitov:1999bv}
J. N. Abdurashitov et al.(SAGE collaboration),
Measurement of the solar neutrino capture rate by SAGE and 
implications  for neutrino oscillations in vacuum",
Phys. Rev. Lett. 83 (1999) 4686-4689; astro-ph/9907131.

\bibitem{Hampel:1998xg}
W. Hampel et al. 
(GALLEX Collaboration),
"GALLEX solar neutrino observations: Results for GALLEX IV",
Phys. Lett. B447 (1999) 127-133.

\bibitem{Ahmad:2001an}
Q. R. Ahmad et al. 
(SNO Collaboration),
"Measurement of the charged current interactions produced by
B-8  solar neutrinos at the Sudbury Neutrino Observatory",
Phys. Rev. Lett. 87 (2001) 071301;
nucl-ex/0106015.

\bibitem{Ahmed:2003kj}
Q. R. Ahmad et al. 
(SNO Collaboration),
"Measurement of the total active B-8 solar neutrino flux at
the Sudbury Neutrino Observatory with enhanced neutral current sensitivity",
Phys. Rev. Lett. 92 (2004) 181301;
nucl-ex/0309004.

\bibitem{Fukuda:2002pe}
S. Fukuda et al. (Super-Kamiokande Collab.),
Determination of solar neutrino oscillation parameters
using 1496 days  of Super-Kamiokande-I data",
Phys. Lett. B539 (2002) 179-187;
hep-ex/0205075.

\bibitem{Apollonio:1999ae}
M. Apollonio et al. 
(CHOOZ Collaboration),
"Limits on neutrino oscillations from the CHOOZ experiment",
Phys. Lett. B466 (1999) 415-430;
hep-ex/9907037.

\bibitem{Boehm:2001ik}
F. Boehm et al.,
"Final results from the Palo Verde neutrino oscillation experiment",
Phys. Rev. D64 (2001) 112001;
hep-ex/0107009.

\bibitem{Eguchi:2002dm}
K. Eguchi et al. 
(KamLAND Collaboration),
"First results from KamLAND: Evidence for reactor anti-neutrino disappearance",
Phys. Rev. Lett. 90 (2003) 021802;
hep-ex/0212021.

\bibitem{Aliu:2004sq}
E. Aliu et al. 
(K2K  Collaboration),
Evidence for muon neutrino oscillation in an accelerator-based experiment",
Phys. Rev. Lett. 94 (2005) 081802;
hep-ex/0411038.

\bibitem{wolfenstein}
L. Wolfenstein, 
Phys. Rev. D17 (1978) 2369.

\bibitem{Aguilar:2001ty}
A. Aguilar et al. 
(LSND Collaboration),
"Evidence for neutrino oscillations from the observation of
anti-nu/e  appearance in a anti-nu/mu beam",
Phys. Rev. D64 (2001) 112007;
hep-ex/0104049.

\bibitem{nova}
I. Ambats et al. 
(NOvA Collaboration),
"NOvA: Proposal to Build an Off-Axis Detector to Study muon-
neutrino $\to$ electron-neutrino Oscillations in the NuMI Beamline",
FERMILAB-PROPOSAL-0929.

\bibitem{Huber:2002mx}
P. Huber, M. Lindner and W. Winter,
"Superbeams versus neutrino factories",
Nucl. Phys. B645 (2002) 3-48;
hep-ph/0204352.

\bibitem{Huber:2004ug}
P. Huber, M. Lindner, M. Rolinec, T. Schwetz,
and W. Winter,
"Prospects of accelerator and reactor neutrino oscillation 
experiments for the coming ten years",
Phys. Rev. D70 (2004) 073014; 
hep-ph/0403068.

\bibitem{nf}
S. Geer,
Phys. Rev. D57 (1998) 6989.

\bibitem{debbie}
D.~A.~Harris et al. 
(Minerva Collaboration),
"Neutrino Scattering Uncertain ties and their Role in Long Baseline Oscillation Experiments",
MINER$\nu$A Note 800, September, 2004;
hep-ex/0410010.

\bibitem{Church:1997ry}
E. Church et al. 
(BooNE Collaboration),
"A proposal for an experiment to measure muon-neutrino $\to$
electron-neutrino oscillations and muon-neutrino
disappearance at the Fermilab Booster: BooNE",
FERMILAB-PROPOSAL-0898.

\bibitem{jlab}
M.K. Jones et al.,
Phys. Rev. Lett. 84 (2000) 1398; 
O. Gayou et al., 
Phys. Rev, Lett. 88 (2002) 092301.

\bibitem{shrock}
R. Shrock, 
Phys. Rev. D12 (1975) 2049; 
A. A. Amer, 
Phys. Rev. D18 (1978) 2290.

\bibitem{brice}
S. J. Brice, S. Geer, K. Paul, and R. Tayloe,
``Low-Energy Neutrino Beams with an Upgraded Fermilab Proton Driver'',
hep-ex/0408135.

\bibitem{prism}
Y. Kuno and Y. Okada, 
Rev. Mod. Phys. 71 (2001) 151-201.

\bibitem{flrp}
``The Coming Revolution in Particle Physics'', 
Report of the Fermilab Long Range Planning Committee, 
May 2004, 
http://www.fnal.gov/pub/today/directors-corner/lrpreportfinal.pdf

\bibitem{Mena:2004}
O. Mena and S. Parke,
Phys. Rev. D69 (2004) 117301.

\end{thebibliography}

\end{document}